\documentclass[a4paper,fleqn,usenatbib]{mnras}

\usepackage{newtxtext,newtxmath}
\usepackage[T1]{fontenc}
\usepackage{ae,aecompl}
\usepackage{graphicx}	% Including figure files
\usepackage{amsmath}	% Advanced maths commands
\usepackage{amssymb}	% Extra maths symbols
\usepackage{float}
\usepackage{subfig}
\usepackage{rotating}

\usepackage{etoolbox}
\makeatletter
\patchcmd\@combinedblfloats{\box\@outputbox}{\unvbox\@outputbox}{}{%
  \errmessage{\noexpand\@combinedblfloats could not be patched}%
}%
\makeatother

\title[Measuring IP mass with \emph{NuSTAR}]{Measuring the masses of Intermediate Polars with \emph{NuSTAR}: V709\,Cas, NY\,Lup and V1223\,Sgr}

\author[A. W. Shaw et al.]{
A. W. Shaw,$^{1}$\thanks{E-mail: aarran@ualberta.ca (AWS)}
C. O. Heinke,$^{1}$
K. Mukai,$^{2,3}$
G. R. Sivakoff,$^{1}$
J. A. Tomsick,$^{4}$
\newauthor and V. Rana$^{5}$
\\
$^{1}$Department of Physics, University of Alberta, CCIS 4-181, Edmonton, AB T6G 2E1, Canada\\
$^{2}$CRESST and X-ray Astrophysics Laboratory, NASA Goddard Space Flight Center, Greenbelt, MD 20771, USA\\
$^{3}$Department of Physics, University of Maryland, Baltimore County, 1000 Hilltop Circle, Baltimore, MD 21250, USA\\
$^{4}$Space Sciences Laboratory, 7 Gauss Way, University of California, Berkeley, CA 94720-7450, USA\\
$^{5}$Raman Research Institute, Sadashivanagar, Bengaluru 560 080, India
}

\date{Accepted XXX. Received YYY; in original form ZZZ}

\pubyear{2017}

\begin{document}
\label{firstpage}
\pagerange{\pageref{firstpage}--\pageref{lastpage}}
\maketitle

\begin{abstract}
The X-ray spectra of intermediate polars can be modelled to give a direct measurement of white dwarf mass. Here we fit accretion column models to \emph{NuSTAR} spectra of three intermediate polars; V709\,Cas, NY\,Lup and V1223\,Sgr in order to determine their masses. From fits to 3--78 keV spectra, we find masses of $M_{\rm WD}=0.88^{+0.05}_{-0.04}M_{\odot}$, $1.16^{+0.04}_{-0.02}M_{\odot}$ and $0.75\pm0.02M_{\odot}$ for V709\,Cas, NY\,Lup and V1223\,Sgr, respectively. Our measurements are generally in agreement with those determined by previous surveys of intermediate polars, but with typically a factor $\sim2$ smaller uncertainties. This work paves the way for an approved \emph{NuSTAR} Legacy Survey of white dwarf masses in intermediate polars.
\end{abstract}


%%%%%%%%%%%%%%%%%%%%%%%%%%%%%%%%%%%%%%%%%%%%%%%%%%

%%%%%%%%%%%%%%%%% BODY OF PAPER %%%%%%%%%%%%%%%%%%

\section{Introduction}
\label{Intro}
Cataclysmic variables (CVs) are binary systems in which a white dwarf (WD) accretes matter, via an accretion disc, from a stellar companion filling its Roche lobe. Magnetic CVs (mCVs) contain WDs with strong magnetic fields ($\sim10^6-10^8$G) which forces material to travel along magnetic field lines. In intermediate polars (IPs), a class of mCVs, the innermost regions of the disc are disrupted within the magnetosphere and in-falling material is instead funnelled on to the poles of the WD in a so-called ``accretion curtain'' \citep[for a review see][]{Patterson-1994}. Close to the surface of the WD, the in-falling material forms a standing shock with a high temperature ($>10$ keV) which emits hard X-rays via optically thin thermal plasma emission, cooling the gas in the post shock region as it descends on to the WD \citep{Aizu-1973}.

The WD mass is a fundamental parameter in quantitative studies of any individual CVs. Moreover, it is important to know the WD mass distribution of CVs as a class for the understanding of their formation and evolution. While accretion alone would act to increase the WD mass in CVs, nova eruptions (thermonuclear runaway of accreted matter) were expected to expel all accreted matter. It was therefore a surprise that WD mass in CVs were found to be higher on average than in pre-CVs \citep{Zorotovic-2011}. However, given the potential implications, it is important to confirm their results using independent methods. Moreover, a comparison of WD mass distributions between non-magnetic and magnetic CVs is of great interest in the context of the origin of the magnetic field in WDs. A leading scenario for the single magnetic WDs is that they are the results of mergers during the common envelope phase; mCVs are then understood to be the consequence of close interaction during the common envelope phase that end just short of merger \citep[][and references therein]{Ferrario-2015}. Such a scenario could lead to a measurable difference between the mean WD masses of magnetic and non-magnetic CVs.

It has been shown that the temperature of the shock in IPs scales with WD mass \citep{Katz-1977, Rothschild-1981} and therefore spectroscopy of IPs in the hard X-ray regime can be used to determine WD masses. It has been noted that X-ray spectroscopy provides a method of measuring WD masses independent of traditional radial velocity studies, which are dominated by uncertainties in binary inclination \citep{Suleimanov-2005,Yuasa-2010}.

X-ray surveys of IPs have led to mass determinations of $\sim20$ IPs \citep{Ramsay-2000,Suleimanov-2005, Brunschweiger-2009, Yuasa-2010}. Unfortunately, systematic uncertainties stemming from poor sensitivities at high energies, and uncertain background \citep[which must be modelled rather than extracted for non-imaging instruments such as \emph{Suzaku's} Hard X-ray Detector;][]{Fukazawa-2009} means that a number of WD masses remain poorly constrained. However, the emergence of the \emph{Nuclear Spectroscopic Telescope Array} (\emph{NuSTAR}) as the first X-ray telescope capable of focusing hard X-rays \citep{Harrison-2013} has brought about the ability to perform high angular resolution spectroscopy beyond 20 keV. This makes \emph{NuSTAR} the ideal observatory to make accurate measurements of WD masses in mCVs. Mass measurements for 3 IPs observed with \emph{NuSTAR} have been published so far \citep{Suleimanov-2016,Hailey-2016}.

In this work we present the results of WD mass measurements of 3 IPs, based on \emph{NuSTAR} spectroscopy. We compare our results to those of earlier studies on these WD masses and note the tighter constraints provided by our mass determinations. This paper, along with previous work, will pave the way for an approved \emph{NuSTAR} Legacy Survey of WD masses in IPs, which will constrain the mass distribution in this particular class of mCVs.

\section{Observations and Data Reduction}

\emph{NuSTAR} consists of two co-aligned focal plane modules, FPMA and FPMB, and is the first telescope in orbit able to focus X-rays up to $\sim79$ keV \citep{Harrison-2013}. \emph{NuSTAR} observed the IPs V709 Cas, NY Lup and V1223 Sgr in 2014 July, August and September, respectively for 20--26ks (Table \ref{NS-exp}). These observations were previously reported by \citet{Mukai-2015}. 

%Table of observations goes here
\begin{table}
	\centering
	\caption{Summary of \emph{NuSTAR} observations}
	\begin{tabular}{l c c c}
	\hline
	Source & Obs. Date & Obs. Time & Exposure (ks) \\
	\hline \hline
	V709\,Cas & 2014-07-07 & 02:01 & 25.6 \\
	NY\,Lup & 2014-08-09 & 14:51 & 23.0 \\
	V1223\,Sgr & 2014-09-16 & 02:26 & 20.4 \\
	\hline
	\end{tabular}
\label{NS-exp}
\end{table}

We reduced the data using the \emph{NuSTAR} data analysis software (NuSTARDAS) v1.8.0, packaged with \textsc{heasoft} v6.22.1. The task {\tt nupipeline} filtered the observations for passages of high background during passages through the South Atlantic Anomaly and produced cleaned event files. For each target and instrument, we extracted spectra from a $70''$ radius circular region centred on the source. Background spectra were extracted from a $100''$ radius circular region centred on a source-free region on the same detector. Spectra and corresponding response matrices were all generated with the task {\tt nuproducts}. Finally, we grouped each spectrum to have at least 25 counts per bin using the \textsc{heasoft} task {\tt grppha}. All spectra were fit in \textsc{xspec} v12.9.1 \citep{Arnaud-1996} using the $\chi^2$ statistic.

\subsection{Modelling IP spectra}

The shocked gas in IPs cools via optically thin thermal plasma emission as it descends on to the WD. The continuum of the overall spectrum in the hard X-ray band can therefore be broadly modelled by a series of bremsstrahlung components. However, since the shock is formed close to the stellar surface, a significant fraction of the radiation will be directed towards the WD and reflected. Reflection manifests itself in the X-ray spectrum as a Compton `hump' at $\sim10-30$ keV and neutral Fe--K emission at $6.4$ keV and has been demonstrated to be extremely important in modelling the X-ray spectrum of IPs \citep{Mukai-2015}. Finally, the immediate pre-shock column have an effect on the X-ray spectrum \citep{Done-1998}, as the emitted X-rays may be viewed through this material some of the time, causing an absorption effect additional to that of the Galactic interstellar medium (ISM). To account for this, some model spectra require the addition of a partial covering component.

For the three IPs discussed in this work, we utilize the IP mass (IPM) model \footnote{IPM is implemented as an additive table model:\newline \citep[{\tt atable\{polarmodel.fits\}}; see e.g.][]{Hailey-2016}} described by \citet{Suleimanov-2005}, which derives a WD mass based on the temperature of the bremsstrahlung continuum, assuming the \citet{Nauenberg-1972} relation between WD mass and radius. The \citet{Suleimanov-2005} model calculates the predicted spectrum by modelling the vertical structure of the accretion column, taking into account the varying gravitational potential over the height of the shock. However, the model does not account for the effects of Compton reflection. We rectify this by including the convolution model for reflection from neutral material {\tt reflect} \citep{Magdziarz-1995}, assuming typical ISM abundances \citep{Wilms-2000}. This introduces the inclination angle of the reflecting surface $\theta$ (characterised in {\tt reflect} by its cosine, $\cos\theta$) as a parameter. However, the derived mass does not depend heavily on $\cos\theta$, in contrast to binary inclination, on which mass has a cubic dependence in radial velocity studies. As the WD mass is primarily determined by the shape of the continuum, we exclude the part of the spectrum in the $5.5-7.5$ keV range, effectively removing the Fe line complex, which has previously been studied in these three sources by \citet{Mukai-2015}. We also include a partial-covering term in some models, using the {\tt pcfabs} model. However, we only use this if we find that a reflection model does not effectively model the spectrum, as partial covering can model some of the X-ray reflection effects, leading to degeneracies between the two models \citep{Yuasa-2010}. In all combinations of models, the Galactic interstellar absorption was accounted for with {\tt tbabs} \citep{Wilms-2000}. The cross-normalisation between the FPMA/B instruments was accounted for by a constant, $C_{\rm FPMB}$.

We first fit the spectra in the energy range 3--78 keV (excluding the Fe line energy range) to derive a mass estimate. However, it has also been shown that similarly robust masses can be obtained by restricting the energy band $\gtrsim15$ keV \citep{Hailey-2016}. The advantage of this method is that absorption and reflection effects can often be ignored, as they have been seen to be negligible at these energies, and we can investigate the effect (if any) they have on the derived masses. 

\section{Results}

\begin{figure*}
\captionsetup[subfigure]{labelformat=empty}
\begin{centering}
\subfloat[]{\includegraphics[width=.45\textwidth]{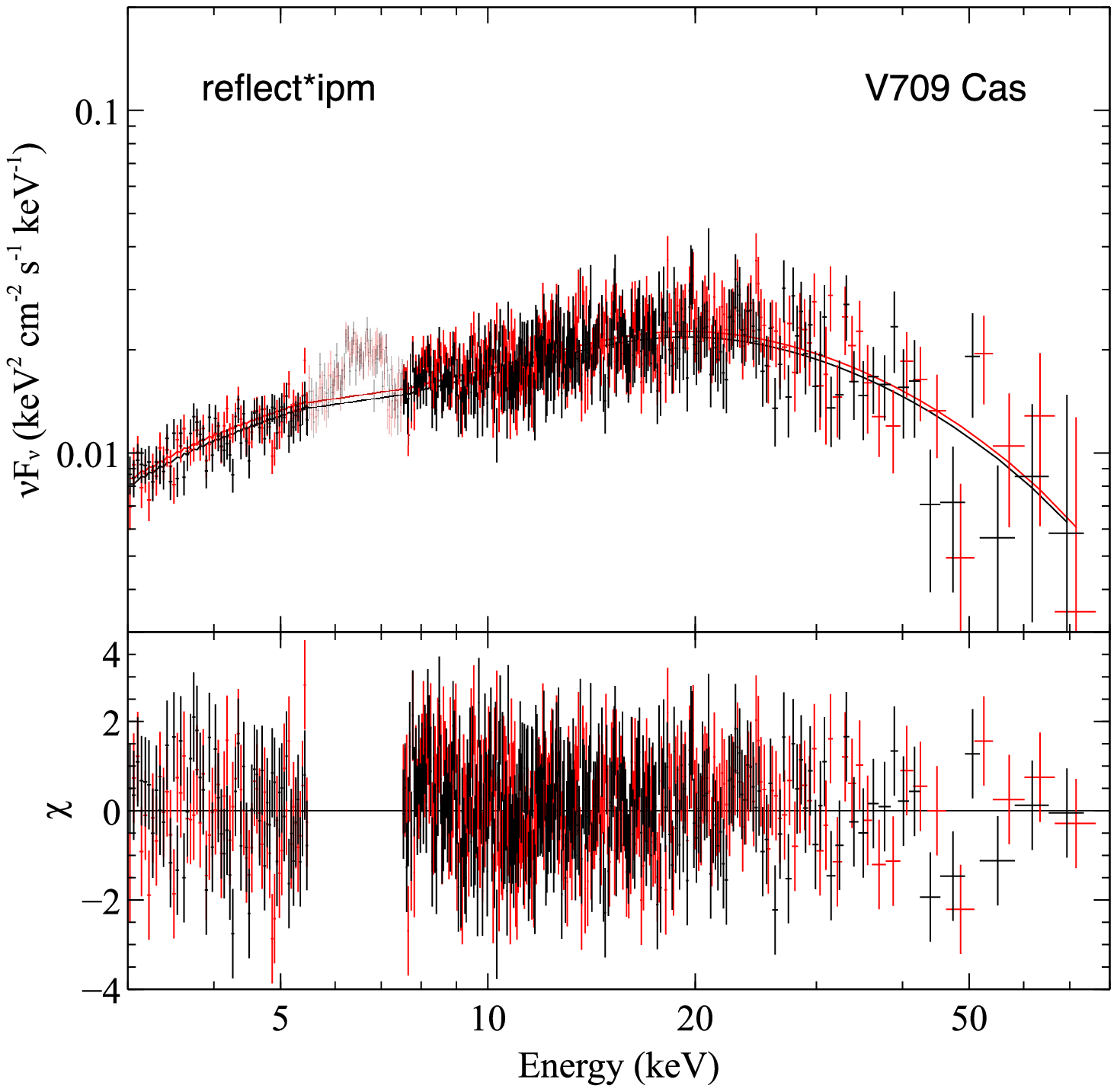}}
\subfloat[]{\includegraphics[width=.45\textwidth]{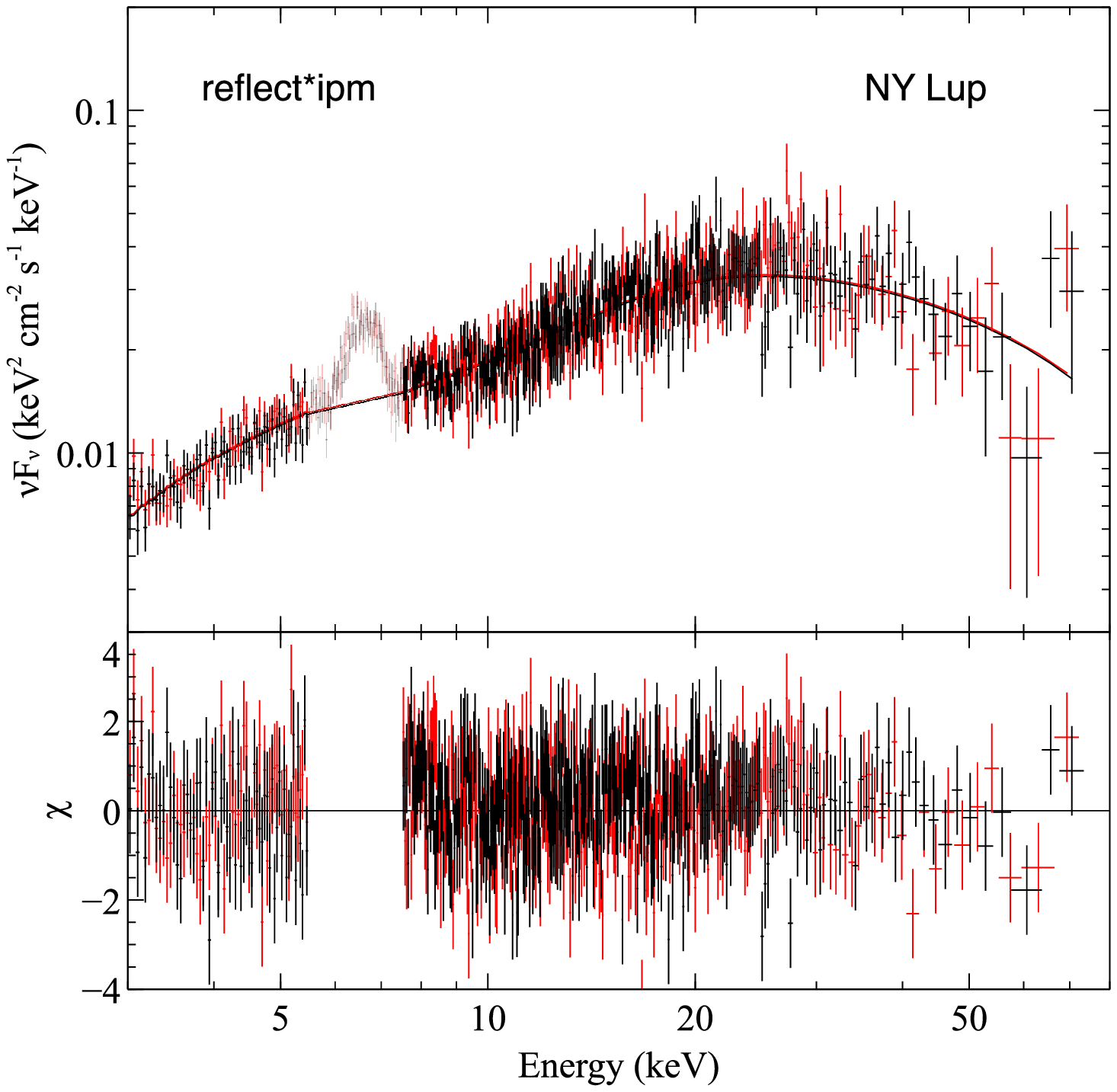}}\\
\vspace*{-2\baselineskip}
\subfloat[]{\includegraphics[width=.45\textwidth]{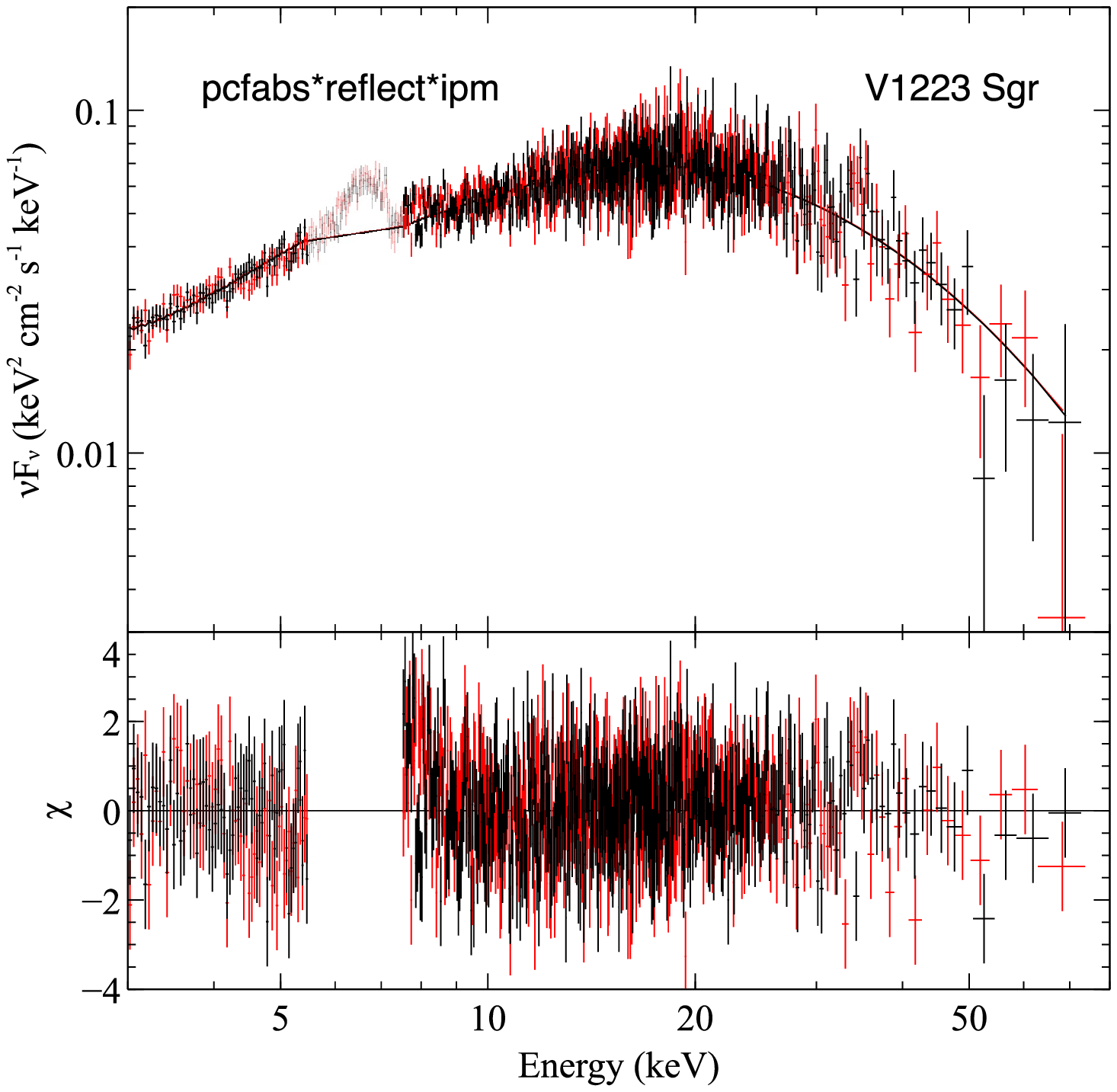}}
\end{centering}
\caption{Unfolded 3--78 keV \emph{NuSTAR} spectra of V709\,Cas, NY\,Lup and V1223\,Sgr, presented in the $\nu F_{\nu}$ regime. Black and red points represent FPMA and FPMB spectra, respectively. The best fit model for each source spectrum is stated in each panel and plotted as a solid line and $\Delta\chi$ residuals are plotted in the bottom panel of each spectrum. We plot the iron emission line complex (5.5--7.5 keV), which we excluded from the fits, with the same colour scheme but a lighter shade.}
\label{spectra}
\end{figure*}

The 3--78 keV spectra and best fit models for each source are shown in Fig. \ref{spectra}. We compare spectral fitting results for different models in Tables \ref{V709-table}, \ref{NY-table} and \ref{V1223-table}.

\subsection{V709\,Cas}

\begin{table*}
\centering
\caption{V709\,Cas: Spectral fitting results}
\label{V709-table}
{
\begin{tabular*}{\linewidth}{@{\extracolsep{\fill}}lccc}
\hline
               \multicolumn{4}{c}{V709\,Cas}                     \\ \hline
Parameters     & {\tt reflect*ipm}$^{\rm a}$ & {\tt reflect*ipm}$^{\rm a}$ & {\tt ipm} \\ \hline\hline
Bandpass [keV] & 3--78$^{\rm b}$             & 3--78$^{\rm b}$             & 15--78    \\ \\
   $N_{\rm H}$ [$10^{22}$ cm$^{-2}$]           &    $3.43\pm0.74$               &        $3.51^{+0.67}_{-0.66}$           & ...          \\
      ${\tt rel_{refl}}$         &    $1.17^{+0.31}_{-0.27}$               &          $1.0^{\rm c}$          &    ...       \\
      $Z^{\rm d}$         &       $1.0$            &       1.0            &     ...      \\
       $Z_{\rm Fe}^{\rm e}$     &         $1.0$            &          1.0         &     ...      \\
       $\cos\theta$        &    $0.45^{\rm c}$               &        $0.62^{+0.24}_{-0.20}$           &     ...      \\
        $M_{\rm WD}$ [$M_{\odot}$]       &        $0.89\pm0.05$           &         $0.88^{+0.05}_{-0.04}$          &     $0.91\pm0.05$     \\
        $N_{\rm IPM}$ [$10^{-11}$]        &     $1.04^{+0.09}_{-0.08}$              &        $1.07\pm0.10$           &     $1.53^{+0.27}_{-0.23}$\\
        $C_{\rm FPMB}$      &         $1.04\pm0.02$          &        $1.04\pm0.02$           &    $1.06\pm0.05$        \\
        Flux [$10^{-11}$ ergs cm$^{-2}$ s$^{-1}$]$^{\rm f}$       &        8.1           &          8.0         &     4.3      \\
        $\chi^2/{\rm dof}$       &    $676/659$               &       675/659            &      212/213  \\
        \hline
\end{tabular*}%
}\\[1ex]
\raggedright $^{\rm a}$\small Multiplied by {\tt tbabs} to account for Galactic absorption\\
\raggedright $^{\rm b}$\small 3--78 keV bandpasses exclude the iron line complex (5.5--7.5 keV)\\
\raggedright $^{\rm c}$\small Fixed\\
\raggedright $^{\rm d}$\small Elemental abundance relative to Solar (fixed)\\
\raggedright $^{\rm e}$\small Iron abundance relative to Solar (fixed)\\
\raggedright $^{\rm f}$\small Observed flux in the bandpass
\end{table*}

V709\,Cas was first identified as an IP from follow-up of a \emph{ROSAT} all-sky survey detection \citep{Haberl-1995,Motch-1996}. The WD has a spin period $P_{\rm spin}=312.8$s \citep{Haberl-1995, Norton-1999} and the orbital period of the binary is 5.33h \citep{Thorstensen-2010}. Modelling the X-ray emission with accretion column models has produced a range of WD mass for V709\,Cas \citep[$0.90-1.22M_{\odot}$;][]{Ramsay-2000,Suleimanov-2005, Brunschweiger-2009, Yuasa-2010}. Neither the WD, nor its companion are detectable in optical spectra \citep{Thorstensen-2010}, so optical radial velocity studies have not been possible. X-ray spectroscopy is therefore the only reliable method of determining the mass of the WD, so we aim here to resolve the prior disagreements with \emph{NuSTAR}.

The IPM model with reflection ({\tt reflect*ipm}) provides a good fit to the 3--78 keV (excluding 5.5--7.5 keV) \emph{NuSTAR} spectra ($\chi^2_{\nu}=1.03$ for 659 degrees of freedom, dof) with a WD mass $M_{{\rm WD}}=0.89\pm0.05 M_{\odot}$. However, the ${\tt rel_{refl}}$ parameter, which characterizes the fraction of downward radiation that is reflected, is greater than unity, which is unphysical. We therefore freeze ${\tt rel_{refl}}$ to unity and instead allow $\cos\theta$ to vary and find a consistent $M_{{\rm WD}}=0.88^{+0.05}_{-0.04} M_{\odot}$ with $\cos\theta=0.62^{+0.24}_{-0.20}$.

A simple IPM model fit to the 15--78 keV spectrum reveals a good $\chi^2_{\nu}=1.00$ (213 dof) and $M_{{\rm WD}}=0.91\pm0.05M_{\odot}$, in agreement with the 3--78 keV band {\tt reflect*ipm} fit. Our measurements of $M_{{\rm WD}}$ are in agreement with the masses of $1.08^{+0.05}_{-0.17}$,  $0.90\pm0.10$ and $0.96\pm0.05M_{\odot}$ derived by \citet{Ramsay-2000}, \citet{Suleimanov-2005} and \citet{Brunschweiger-2009}, respectively. However, our measurements do not agree with the $M_{{\rm WD}}=1.22^{+0.05}_{-0.20}M_{\odot}$ as derived by \citet{Yuasa-2010}.

\subsection{NY\,Lup}

\begin{table*}
\centering
\caption{NY\,Lup: Spectral fitting results}
\label{NY-table}
{
\begin{tabular*}{\linewidth}{@{\extracolsep{\fill}}lcccc}
\hline
                \multicolumn{5}{c}{NY\,Lup}                     \\ \hline
Parameters     & {\tt reflect*ipm}$^{\rm a}$ & {\tt reflect*ipm}$^{\rm a}$ & {\tt ipm} & {\tt reflect*ipm} \\ \hline\hline
Bandpass (keV) & 3--78$^{\rm b}$             & 3--78$^{\rm b}$             & 15--78  & 15--78   \\ \\ 
 $N_{\rm H}$ [$10^{22}$ cm$^{-2}$]           &    $4.03\pm0.83$               &        $4.75^{+0.70}_{-0.64}$           & ...           & ... \\
       ${\tt rel_{refl}}$         &    $1.76^{+0.33}_{-0.29}$               &          1.0$^{\rm c}$         &    ...    & 1.0$^{\rm c}$   \\
      $Z^{\rm c}$         &       1.0            &       1.0            &     ...    & 1.0 \\
       $Z_{\rm Fe}^{\rm c}$     &         $1.0$            &          1.0         &     ...    & 1.0  \\
       $\cos\theta$        &    $0.45^{\rm c}$               &        $0.94_{-0.16}$           &     ...   &  $0.95_{-0.24}$ \\
        $M_{\rm WD}$ [$M_{\odot}$]       &        $1.15\pm0.05$           &         $1.16^{+0.04}_{-0.02}$          &     $1.20\pm0.05$  & $1.17\pm0.05$  \\
        $N_{\rm IPM}$ [$10^{-11}$]        &     $0.57\pm0.05$              &        $0.59^{+0.03}_{-0.05}$           &     $0.96^{+0.16}_{-0.14}$ & $0.58^{+0.10}_{-0.03}$\\
        $C_{\rm FPMB}$      &         $1.01\pm0.02$          &        $1.01\pm0.02$           &    $0.98\pm0.04$  &   $0.98\pm0.04$   \\
        Flux [$10^{-11}$ ergs cm$^{-2}$ s$^{-1}$]$^{\rm d}$       &        11.1           &          11.1         &  7.6      &  7.3 \\
        $\chi^2/{\rm dof}$       &    $730/724$               &       725/724            &      307/270        & 269/269\\
        \hline
\end{tabular*}%
}\\[1ex]
\raggedright $^{\rm a}$\small Multiplied by {\tt tbabs} to account for Galactic absorption\\
\raggedright $^{\rm b}$\small 3--78 keV bandpasses exclude the iron line complex (5.5--7.5 keV)\\
\raggedright $^{\rm c}$\small Fixed\\
\raggedright $^{\rm d}$\small Observed flux in the bandpass\\
\end{table*}

NY\,Lup was identified as an IP from follow-up of the \emph{ROSAT} all-sky survey \citep{Haberl-2002}. The WD has $P_{\rm spin}=693$s and a binary orbital period of 9.87h \citep{Haberl-2002,deMartino-2006}. A radial velocity study of the mass donor suggested $M_{{\rm WD}}\geq0.5M_{\odot}$ \citep{deMartino-2006}, in agreement with X-ray spectral modelling \citep[$M_{{\rm WD}}=1.09\pm0.07 M_{\odot}$, $M_{{\rm WD}}=1.15^{+0.08}_{-0.07} M_{\odot}$;][respectively]{Brunschweiger-2009,Yuasa-2010}. NY\,Lup shows colour-dependent circular polarization modulated on the spin period of the WD indicative of a fairly strong ($>4$ MG) magnetic field \citep{Katajainen-2010,Potter-2012}.

As in the case of V709\,Cas, we obtain a good fit ($\chi^2_{\nu}=1.01$; 724 dof) with an IPM model with reflection to the 3--78 keV spectra, but with an unphysical ${\tt rel_{refl}}>1$. To rectify this, we freeze ${\tt rel_{refl}}=1$, allowing $\cos\theta$ to vary, and find a best fit ($\chi^2_{\nu}=1.00$; 724 dof) $M_{{\rm WD}}=1.16^{+0.04}_{-0.02} M_{\odot}$. In this case we find that $\cos\theta$ tends toward the hard upper limit of 0.95, with the best fit value $\cos\theta\geq0.78$ (90\% confidence). This is consistent with the low value for $\theta$ favoured by \citet{Mukai-2015}.

Fitting an IPM model, without reflection, to the 15--78 keV band spectrum indicates a heavier WD ($M_{{\rm WD}}=1.20\pm0.05 M_{\odot}$), though still consistent with the 3--78 keV band fit within uncertainties. However, it is worth noting that the 15--78 keV fit has a higher $\chi^2_{\nu}=1.14$ (270 dof). This is not surprising as it has been shown that NY\,Lup has the highest amplitude Compton reflection hump, in addition to the strongest iron line (equivalent width = $132\pm12$ eV) of the three sources studied here \citep{Mukai-2015}, so reflection is still likely a significant contribution to the shape of the X-ray spectrum, even at these energies. Convolving the IPM model with the \citet{Magdziarz-1995} reflection model in the 15--78 keV band improves the fit ($\Delta\chi^2 = 38$ for one fewer dof; $\chi^2_{\nu}=0.99$) and we find $M_{{\rm WD}}=1.17\pm0.05 M_{\odot}$, consistent with the full 3--78 keV band fit. The masses we derive for NY\,Lup are in good agreement with the masses determined by \citet{Brunschweiger-2009} and \citet{Yuasa-2010} and it is evident that the system contains a relatively heavy WD.

\subsection{V1223\,Sgr}

\begin{table*}
\begin{center}
\caption{V1223\,Sgr: Spectral fitting results}
\label{V1223-table}
\begin{tabular*}{\linewidth}{@{\extracolsep{\fill}}lcccc}
\hline
                \multicolumn{5}{c}{V1223\,Sgr}                     \\ \hline
Parameters     & {\tt reflect*ipm}$^{\rm a}$ & {\tt reflect*ipm}$^{\rm a}$ & {\tt reflect*pcfabs*ipm}$^{\rm a}$ & {\tt ipm} \\ \hline\hline
Bandpass (keV) & 3--78$^{\rm b}$             & 3--78$^{\rm b}$             & 3--78$^{\rm b}$  & 15--78   \\ \\
$N_{\rm H}$ [$10^{22}$ cm$^{-2}$]           &    $6.59^{+0.58}_{-0.59}$               &        $7.65^{+0.49}_{-0.30}$           & $0^{\rm c}$           & ... \\
$N^{\rm pc}_{\rm H}$ [$10^{22}$ cm$^{-2}$] & ... & ... & $38.7^{+14.6}_{-5.8}$ & ... \\
 $f^{\rm pc}$ & ... & ... & $0.52^{+0.02}_{-0.11}$ & ... \\
      ${\tt rel_{refl}}$         &    $2.06^{+0.29}_{-0.27}$               &          1.0$^{\rm d}$         &    1.0$^{\rm d}$   & ...    \\
      $Z^{\rm d}$         &       $1.0$            &       1.0            &     1.0    & ... \\
       $Z_{\rm Fe}^{\rm d}$     &         $1.0$            &          1.0         &     1.0    & ...  \\
       $\cos\theta$        &    $0.45^{\rm d}$               &        $0.95_{-0.10}$           &    $0.52^{+0.18}_{-0.15}$  &  ... \\
        $M_{\rm WD}$ [$M_{\odot}$]       &        $0.75\pm0.02$           &         $0.78\pm0.01$          &     $0.75\pm0.02$  & $0.77\pm0.02$  \\
        $N_{\rm IPM}$ [$10^{-11}$]        &     $4.35\pm0.18$              &        $4.43^{+0.13}_{-0.17}$           &     $5.57^{+0.44}_{-0.24}$ & $7.66^{+0.75}_{-0.68}$\\
        $C_{\rm FPMB}$      &         $1.00\pm0.01$          &        $1.00\pm0.01$           &    $1.00\pm0.01$  &   $1.00\pm0.03$   \\
        Flux [$10^{-11}$ ergs cm$^{-2}$ s$^{-1}$] $^{\rm e}$       &       23.2          &          23.3         &     23.3   &  11.8 \\
        $\chi^2/{\rm dof}$       &    $1006/919$               &       $1027/919$            &      $938/917$       & $417/423$\\
        \hline
\end{tabular*}\\[1ex]
\raggedright $^{\rm a}$\small Multiplied by {\tt tbabs} to account for Galactic absorption\\
\raggedright $^{\rm b}$\small 3--78 keV bandpasses exclude the iron line complex (5.5--7.5 keV)\\
\raggedright $^{\rm c}$\small Insensitive to fit\\
\raggedright $^{\rm d}$\small Fixed\\
\raggedright $^{\rm e}$\small Observed flux in the bandpass
\end{center}
\end{table*}

V1223\,Sgr was discovered by \emph{Uhuru} \citep{Forman-1978}, but only identified as a CV three years later \citep{Steiner-1981}. The WD has $P_{\rm spin}=745.6$s \citep{Osborne-1985} and the binary orbital period is 3.37h \citep{Jablonski-1987}. As one of the brightest and most well studied IPs, there are a range of mass estimates from $0.71-1.05M_{\odot}$ for the WD in V1223\,Sgr \citep[see][and references therein]{Hayashi-2011}.

Neither the {\tt reflect*ipm} with fixed inclination (free reflection fraction), nor with varying inclination (${\tt rel_{refl}}=1$) provide a formally acceptable fit in the 3--78 keV band ($\chi^2_{\nu}=1.10$; 919 dof and $1.12$; 919 dof, respectively). The residuals suggest that the fit is poor below $<5$ keV, indicating that we are likely accounting for absorption incorrectly. To this end, we modify the fit with the addition of a partial covering model ({\tt pcfabs}), to account for absorption by the pre-shock flow. The fit improves ($\chi^2_{\nu}=1.02$; 917 dof) with a partial covering column density $N^{{\rm pc}}_{{\rm H}}=38.7^{+14.6}_{-5.8}\times10^{22} {\rm cm}^{-2}$ and covering fraction $f^{{\rm pc}}=0.52^{+0.02}_{-0.11}$ and we find $M_{{\rm WD}}=0.75\pm0.02 M_{\odot}$ in this case.

The uncovered IPM model, with no reflection, provides a good fit ($\chi^2_{\nu}=0.99$; 423 dof) to the 15--78 keV band spectrum and we find  $M_{{\rm WD}}=0.77\pm0.02 M_{\odot}$, consistent with the 3--78 keV band measurement. Our $M_{{\rm WD}}$ measurements are more indicative of a moderate mass WD, in agreement with previous measurements of $0.93\pm0.12$, $0.75\pm0.05$ and $0.82^{+0.05}_{-0.06}M_{\odot}$ \citep{Beuermann-2004,Yuasa-2010,Hayashi-2011}, rather than a heavy ($>1M_{\odot}$) WD, as suggested by \citet{Ramsay-2000} and \citet{Evans-2007b}.

\section{Discussion}

\subsection{WD mass measurements}
We have shown that it is possible to obtain accurate mass estimates for WDs in IPs by modelling the X-ray spectrum in the 3--78 keV band as a one-dimensional accretion flow, which cools via bremsstrahlung as it descends on to the surface of the WD \citep{Suleimanov-2005}. Though previous studies have suggested that reflection may have a negligible effect on the measured mass \citep{Cropper-1998}, it has been shown that it is an important component in observed X-ray spectra \citep{Mukai-2015}. In the case of V1223\,Sgr, we require a partial covering component to account for the absorption effects of the accretion curtain.

We also constrain consistent WD masses with only the 15--78 keV portion of the spectrum \citep[see also][]{Yuasa-2012,Hailey-2016}. This has the advantage of reducing computational time but still producing a consistent mass estimate, though with slightly larger statistical uncertainties. Still, this method can be used as a robust estimate of WD mass, before more complicated models are considered over the full band. 

The derived masses for the three IPs studied here are mostly consistent with previous measurements. However, the largest discrepancy lies with V709\,Cas, where the mass of the WD was derived to be $1.22^{+0.05}_{-0.20}M_{\odot}$ by \citet{Yuasa-2010}, compared to $M_{{\rm WD}}=0.88^{+0.05}_{-0.04} M_{\odot}$ derived in this work \citep[consistent with][]{Ramsay-2000,Suleimanov-2005,Brunschweiger-2009}. \citet{Yuasa-2010} note that there may be some systematic uncertainties present in their results as a result of the limitations of their model, and we note here the large residuals in the \emph{Suzaku} spectral fits, and the resultant large negative uncertainty in the derived mass. This does not necessarily mean that our work is free from potential systematic uncertainties - we do use the same model as \citet{Suleimanov-2005} and \citet{Brunschweiger-2009} and derive consistent masses - but \emph{NuSTAR} is less susceptible to an uncertain background estimation due to its capability as an imaging telescope. Also, \citet{Yuasa-2010} do not provide a treatment of reflection in their modelling, despite it having been shown to be an important component of the X-ray spectrum of V709\,Cas \citep{Mukai-2015}. We investigate this by fitting an IPM model with no reflection component to the 3--78 keV (excluding 5.5--7.5 keV as before) spectra of V709\,Cas. We find an unacceptable fit ($\chi^2_{\nu}=1.15$; 660 dof) and $M_{{\rm WD}}=1.15\pm0.03 M_{\odot}$, consistent with \citet{Yuasa-2010}. It is therefore evident that the mistreatment of reflection in X-ray spectral fitting can lead to a misinterpretation of WD mass.

\subsection{Underestimation of WD mass?}

The IPM model \citep{Suleimanov-2005} utilizes the link between the temperature of the standing shock and the mass of the WD. The model assumes that the inner radius of the accretion disc, $R_{\rm in}$, which terminates at the magnetospheric radius of the WD, $R_{\rm m}$, is far enough from the WD surface that accreting material can be considered to be free-falling from infinity. However, if $R_{\rm m}$ is sufficiently small ($\lesssim5$ WD radii, $R_{\rm WD}$), then the accretion flow will be accelerated to lower velocities, and the shock will have a lower temperature than derived by the assumed model \citep{Suleimanov-2016}. This leads to the possibility of some WD masses being underestimated by IP mass models.

\begin{figure*}
\captionsetup[subfigure]{labelformat=empty}
\begin{centering}
\subfloat[]{\includegraphics[width=.45\textwidth]{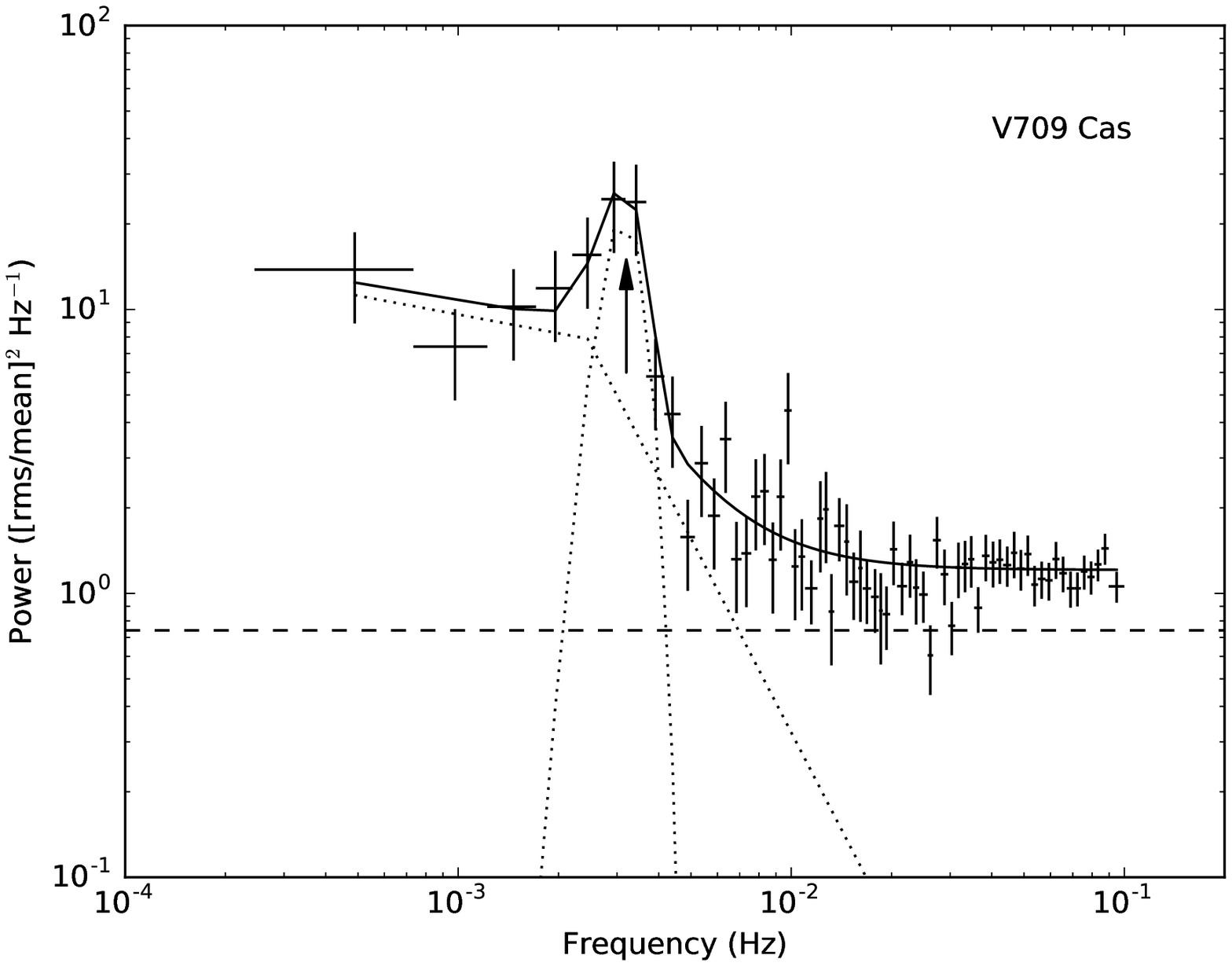}}
\subfloat[]{\includegraphics[width=.45\textwidth]{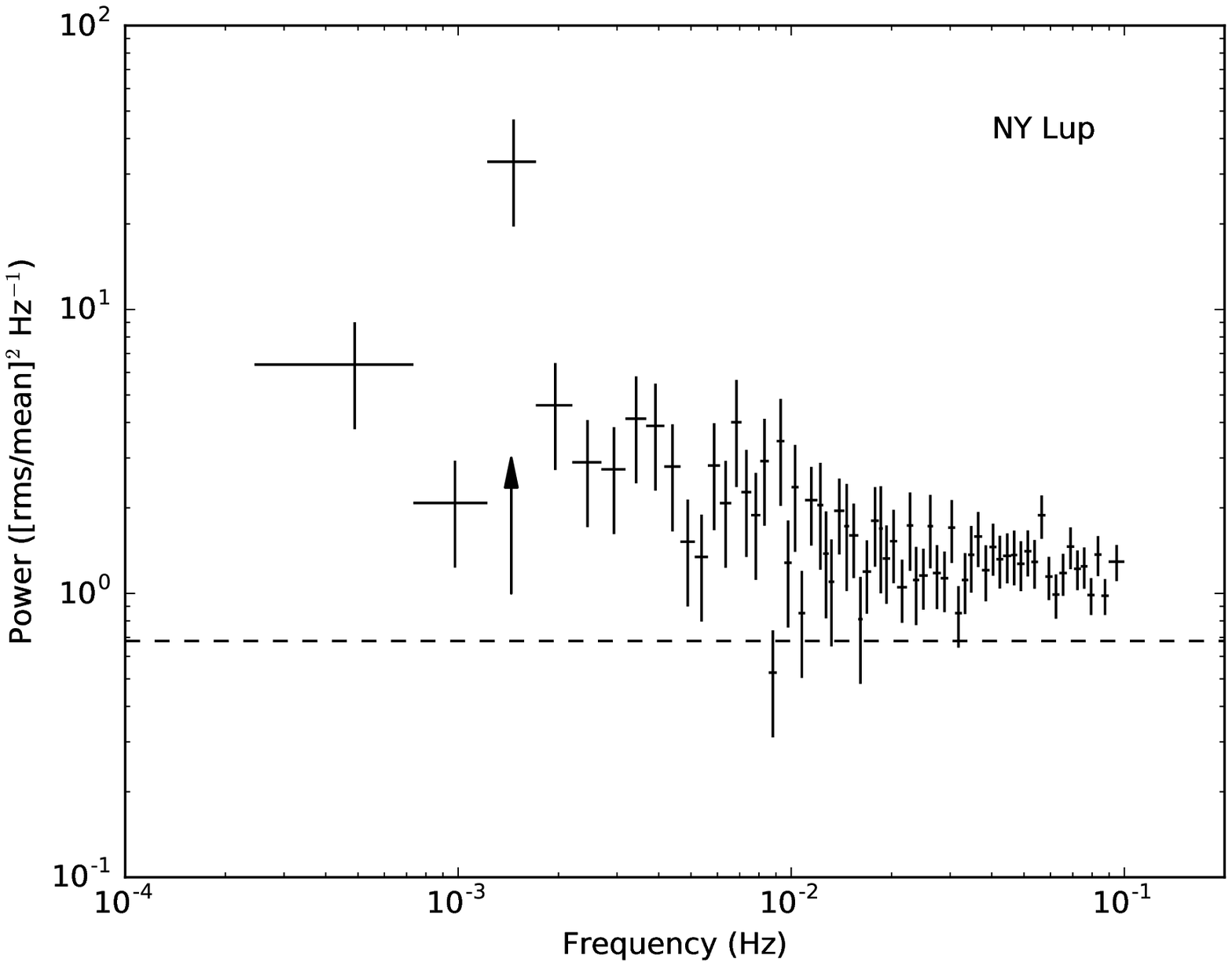}}\\
\vspace*{-3\baselineskip}
\subfloat[]{\includegraphics[width=.45\textwidth]{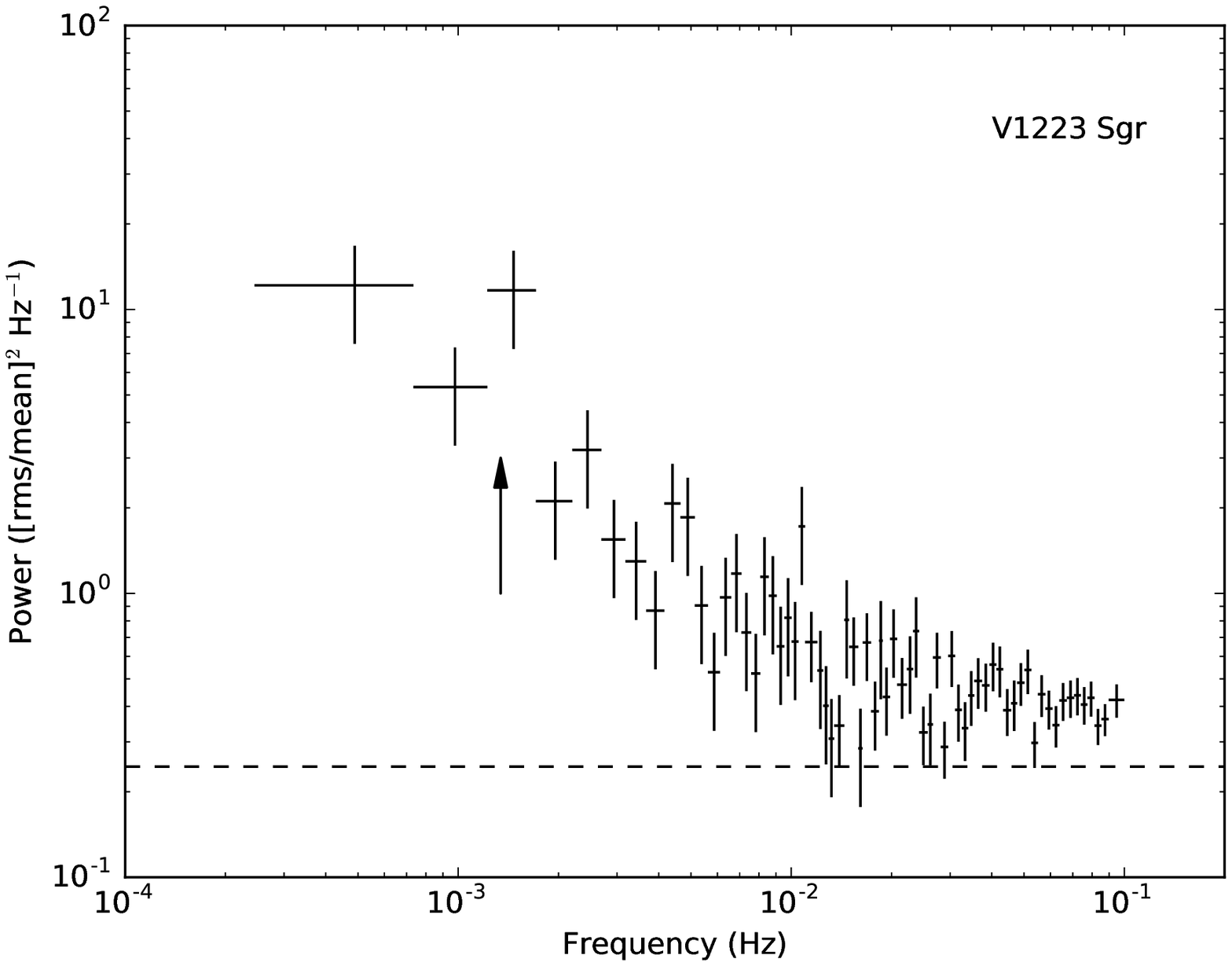}}
\end{centering}
\caption{Power spectra of the X-ray light curves of V709\,Cas, NY\,Lup and V1223\,Sgr. Power spectra are rms normalised \citep{vanderKlis-1997}. The spin periods of the WDs are marked with an arrow and the predicted noise level is denoted by a dashed line (note that this does not take into account detector dead-time). For V709\,Cas, we plot the best fit model consisting of a broken power law with a Gaussian and a $0^{\rm th}$ order polynomial, the Gaussian and the broken power law (with a break at $\nu_{\rm b}=2.4\pm1.0\times10^{-3}$ Hz) components of the best fit model are represented by the dotted lines.}
\label{powerspectra}
\end{figure*}

\citet{Revnivtsev-2009,Revnivtsev-2011} postulated that the frequency of the break that is commonly seen in the power spectra of systems with magnetospherically truncated accretion discs corresponds to the Kepler frequency at $R_{\rm m}$. Therefore, measuring this break could provide an estimate of $R_{\rm m}$. Based on this relation \citet{Suleimanov-2016} claimed that the magnetospheres of the two IPs, GK Per and EX Hya, reached to $<3R_{\rm WD}$, and thus these WD masses had been underestimated previously \citep[e.g. by][]{Suleimanov-2005, Yuasa-2010}.

Here we investigate the possibility of a misinterpretation of $R_{\rm m}$ affecting our mass measurements of the three IPs studied in this work. We construct the power spectra for these three IPs (Fig. \ref{powerspectra}) using the \textsc{python} package MaLTPyNT \citep{Bachetti-2015}. Power spectra are constructed from 2048s light curve segments and combined to produce the averaged power spectrum for each source. We find that NY\,Lup and V1223\,Sgr exhibit no obvious break in their power spectra, with the only visible indicator of discrete variability being the spin period of the WDs in each system. This is contrary to \citet{Revnivtsev-2011} who report a break at $\nu_{\rm b}=0.07$ Hz in optical photometry of V1223\,Sgr. We must note here that white noise dominates the power spectra at frequencies above $\sim0.01$ Hz, so a break at 0.07 Hz would likely be undetectable. However, if the break as measured by \citet{Revnivtsev-2011} is indeed indicative of the Kepler frequency, then this would indicate that $R_{\rm m}$ is extremely close to the WD surface ($\sim1.1R_{\rm WD}$ for a $0.78M_{\odot}$ WD). If this were the case we would see a very soft X-ray spectrum as the accretion flow would not be accelerated to very high velocities before reaching the shock - if accelerated at all - and the post-shock region would have a very low temperature. This is not in agreement with the observed properties of V1223\,Sgr, which exhibits a shock temperature of $kT_{\rm shock}>30$keV \citep{Yuasa-2010,Mukai-2015}, typical of IPs.

The power spectrum of V709\,Cas, however, may show a break close to or at the spin frequency. To investigate, we fit a model consisting of a broken power law, with a Gaussian at the spin frequency, and a $0^{\rm th}$ order polynomial for the white noise at higher frequencies. We find that a power law with a break at $\nu_{\rm b}=2.4\pm1.0\times10^{-3}$ Hz can constrain the power spectrum. However, the fit is indistinguishable at 90\% confidence from a single power law model, with a Gaussian at the spin period ($\Delta\chi^2=2.7$ for one more degree of freedom). We therefore cannot confidently claim the detection of a break and do not attempt to use $\nu_{\rm b}$ to modify the masses we have derived in this work.

We can use the fact that most IPs are approximately in spin equilibrium \citep{King-1991,King-1993} as an independent approximation of $R_{\rm m}$. For mCVs in spin equilibrium, $R_{\rm in}$ (and therefore $R_{\rm m}$) is approximately equal to the co-rotation radius $R_{\rm co}=(GM_{\rm WD}P_{\rm spin}^2/4\pi^2)^{1/3}$ \citep{King-1991}. For NY\,Lup and V\,1223 Sgr $R_{\rm m}>>10R_{\rm WD}$, which is far enough away from the WD surface to have a negligible effect on the shock temperature. However, the low $P_{\rm spin}=312.78$s \citep{Haberl-1995, Norton-1999} of V709\,Cas suggests that $R_{\rm m}$ may be closer to the WD surface. Using the \citet{Nauenberg-1972} relation between WD mass and radius we find $R_{\rm m}\sim10.5R_{\rm WD}$ for a $0.88M_{\odot}$ WD. This is still not close enough to the WD to have a large effect on the derived $M_{\rm WD}$ according to \citet{Suleimanov-2016}. The temperature of the shock $kT_{\rm shock}=\frac{3}{8}\mu m_{\rm H}\frac{GM_{\rm WD}}{R_{\rm WD}}(1-R_{\rm m}/R_{\rm WD})$, where $\mu=0.615$ is the mean molecular weight for a completely ionized plasma with solar abundances and $m_{\rm H}$ is the proton mass. The best-fit mass $M_{\rm WD}=0.88^{+0.05}_{-0.04}M_{\odot}$ implies $kT_{\rm shock}=44.1^{+5.4}_{-3.9}$keV. If we re-calculate the mass using this temperature and instead place the inner disk at $R_{\rm m}=10.5R_{\rm WD}$, we find $M_{\rm WD}=0.92^{+0.05}_{-0.04}M_{\odot}$ - an increase but still consistent within uncertainties with the mass derived from an infinite fall height.

\subsubsection{Location of the standing shock}

Another caveat of the IPM model \citep{Suleimanov-2005} is that it assumes that the standing shock is formed immediately above the WD surface. However, this may not always be the case. To explain the differing amplitudes of reflected emission seen in the three systems studied in this work, \citet{Mukai-2015} propose that the shock may form at a non-negligible distance from the WD surface, dependent on the level of reflection exhibited. V709\,Cas exhibits a low reflection amplitude and thus the shock is likely $\sim0.2R_{\rm WD}$ above the WD. Conversely, NY\,Lup, which shows strong reflection in its spectrum, likely has a shock close to the WD surface, consistent with the assumptions in the \citet{Suleimanov-2005} model. To explain the level of reflection seen in the spectrum of V1223\,Sgr, \citet{Mukai-2015} suggest a shock height of $\sim0.05R_{\rm WD}$.

Underestimating the shock height $R_{\rm shock}$ leads to an underestimation of $M_{\rm WD}$ as it introduces an additional term to the calculation of the shock temperature: $kT_{\rm shock}=\frac{3}{8}\mu m_{\rm H}\frac{GM_{\rm WD}}{(R_{\rm WD}+R_{\rm shock})}(1-R_{\rm m}/R_{\rm WD})$. In the case of V709\,Cas, a shock with temperature $kT_{\rm shock}=44.1^{+5.4}_{-3.9}$keV located $0.2R_{\rm WD}$ above the WD would imply $M_{\rm WD}=0.96\pm0.04M{\odot}$. The inclusion of the fall height at $R_{\rm m}\sim10.5R_{\rm WD}$ would increase the WD mass further to $M_{\rm WD}=1.00^{+0.05}_{-0.04}M_{\odot}$. 

In addition to the case of V709\,Cas, \citet{Mukai-2015} suggest a shock height of $0.05R_{\rm WD}$ for V1223\,Sgr. This would imply $M_{\rm WD}=0.77\pm0.02M_{\odot}$ - a slight increase but consistent within uncertainties with the best-fit mass derived from the IPM model. It is evident that the shock height can have an non-negligible effect on the derived WD mass if it is more than a few tenths of a WD radius above the stellar surface and should be carefully considered in the analysis of IP spectroscopy.

\subsection{Comparison with cooling flow models}

In their study of reflection from the three IPs presented in this work, \citet{Mukai-2015} utilize an isobaric cooling flow model to characterise the emission from the post-shock region \citep[{\tt mkcflow};][]{Mushotzky-1988}. We can use the best-fit shock temperatures to derive the implied mass and compare to the IPM model fits in this work. Assuming the material is falling from infinity and forming a shock just above the surface of the WD \citep[which we note above may not be a completely valid assumption but mirrors the model of][]{Suleimanov-2005}, then $kT_{\rm shock}=\frac{3}{8}\mu m_{\rm H}\frac{GM_{\rm WD}}{R_{\rm WD}}$.

The best fit temperatures are $kT_{\rm shock}=50.0^{+4.6}_{-3.9}$, $55.5^{+6.9}_{-3.4}$ and $35.4^{+1.7}_{-1.5}$ keV for V709\,Cas, NY\,Lup and V1223\,Sgr, respectively. This translates to $M_{\rm WD}=0.94^{+0.04}_{-0.03}$, $0.99^{+0.05}_{-0.03}$ and $0.79\pm0.02M_{\odot}$, respectively, utilising the \citet{Nauenberg-1972} mass-radius relation. For V709\,Cas and V1223\,Sgr, these values are broadly consistent, within uncertainties, with those calculated from our IPM model fits. However, there is a remarkable difference for NY\,Lup, for which we derive $M_{\rm WD}=1.16^{+0.04}_{-0.02}M_{\odot}$ using the IPM model to fit the spectrum. It is worth noting here that, out of the three sources studied in this work, NY\,Lup is the one that is most affected by a reflection component, which may have an effect on either the temperature derived from the cooling flow model, or on the mass derived from the IPM model. However, the likely reason for the discrepancy is the high value of $kT_{\rm shock}>50$ keV, which drives the location of the cut-off in a bremsstrahlung energy spectrum. The cut-off for NY\,Lup is beyond 50 keV, where the background is much more dominant, and thus it is difficult to constrain $kT_{\rm shock}$ tightly. It is possible that the current data do not provide robust constraints on high mass WDs, and that longer exposures are required in order to increase signal-to-noise at high energies.

\subsection{Iron abundance}

In all of our fits, we have assumed \citet{Wilms-2000} solar abundances for the reflecting surface, including of iron. Conversely, analysis of Fe lines in IP spectra suggests that the majority of WDs in IPs have significantly sub-solar Fe abundances \citep[see e.g.][]{Ezuka-1999,Yuasa-2010}. However, as the majority of confirmed IPs are at distances $<500$ pc \citep[][and references therein]{Suleimanov-2005}, and therefore in the solar neighbourhood, we would expect the abundances of the companion stars (and therefore the accreted material) to be similar to that of the Sun, on average. It is unclear if we would expect the abundance of the reflector (i.e. the WD surface) to be significantly sub-solar, but as it is continuously accreting stellar material from a main sequence companion in the solar neighbourhood, then, unless the elements quickly stratify, then it would not be unreasonable to assume solar abundances. We therefore approach the subject of iron abundance with some caution. 

Nevertheless, we show here how changes in abundances can affect reflection and therefore the derived mass. Fixing $Z_{\rm Fe}$ to a value of 0.5 (relative to solar), we find $M_{\rm WD}=0.92\pm0.05$, $1.20\pm0.04$ and $0.78\pm0.02M_{\odot}$ for V709\,Cas, NY\,Lup and V1223\,Sgr, respectively. Lowering iron abundances has the effect of increasing the derived mass (by approximately one error bar), but it is unclear if a sub-solar $Z_{\rm Fe}$ is the correct approach.

We must note here that a degeneracy is present if ${\tt rel_{refl}}$, $\cos\theta$ and $Z_{\rm Fe}$ are all allowed to vary at the same time, so an accurate interpretation of all three parameters simultaneously is not possible with the current modelling strategy. A detailed study of abundances is possible with grating spectroscopy of the iron line region with e.g. \emph{Chandra}/High Energy Transmission Grating Spectrometer (HETGS) or the micro-calorimeter of the future \emph{X-ray Astronomy Recovery Mission} (\emph{XARM}).

\subsection{Toward a systematic IP survey}

Here we have shown that we can accurately model WD masses in IPs (within $\sim5\%$) using relatively short exposures with \emph{NuSTAR}. We can improve on the precision of previous measurements determined using non-imaging telescopes in significantly less time. For example, a 46.2 ks observation of V1223\,Sgr with \emph{Suzaku} enabled \citet{Yuasa-2010} to derive $M_{{\rm WD}}=0.75\pm0.05 M_{\odot}$, compared to the 20.4 ks \emph{NuSTAR} observation presented here from which we calculated $M_{{\rm WD}}=0.75\pm0.02 M_{\odot}$. We even derive tighter constraints from our \emph{NuSTAR} spectra than from 2.5 years of \emph{Swift}/BAT observations \citep{Brunschweiger-2009}. Also, \emph{NuSTAR} is the first telescope to unambiguously detect Compton reflection in IPs \citep{Mukai-2015}, which is an important component in modelling their hard X-ray spectra.

This brings into context the potential efficiency of a systematic survey of IPs with \emph{NuSTAR}. We have therefore devised a $\sim1$Ms \emph{NuSTAR} Legacy Survey\footnote{https://www.nustar.caltech.edu/page/legacy\_surveys\#g7} (PI: Shaw) of 25 mCVs in order to efficiently measure the WD mass distribution in IPs, which has many important implications for binary evolution models and the production of type Ia supernovae \citep{Zorotovic-2011}. To devise the target list we selected the 25 brightest mCVs in the \emph{Swift}/BAT 70 month catalogue \citep{Baumgartner-2013} that have not previously been observed with \emph{NuSTAR}.

\section{Conclusions}
We have modelled the mass of three bright IPs observed with \emph{NuSTAR} and found that the measured masses are consistent with the majority of previously published measurements but with tighter constraints. For NY\,Lup we find a very strong reflection component that contributes to the X-ray spectrum even above 15 keV, confirming the results of \citet{Mukai-2015}. We find no strong evidence for a break in the power spectra of the sources, suggesting that the inner edge of the accretion disc is likely relatively distant from the WD, unlike in GK Per \citep{Suleimanov-2016}. However, we do present our results with caveats, showing that just a slight change in the height of the shock above the WD, the location of the inner accretion disc, or the iron abundance can alter our interpretation of the derived WD mass. This work, along with that of \citet{Hailey-2016} and \citet{Suleimanov-2016} has shown that \emph{NuSTAR} is the ideal facility with which to perform a systematic survey of IPs in order to accurately determine the mass distribution of WDs in mCVs. This will be realised with an approved $\sim1$ Ms \emph{NuSTAR} Legacy Survey.

\section*{Acknowledgements}
We thank the anonymous referee for useful comments which helped improve the manuscript. COH and GRS are supported by NSERC Discovery Grants and COH by a Discovery Accelerator Supplement. This work made use of data from the {\it NuSTAR} mission, a project led by the California Institute of Technology, managed by the Jet Propulsion Laboratory, and funded by the National Aeronautics and Space Administration. We thank the {\it NuSTAR} Operations, Software and  Calibration teams for support with the execution and analysis of these observations.  This research has made use of the {\it NuSTAR}  Data Analysis Software (NuSTARDAS) jointly developed by the ASI Science Data Center (ASDC, Italy) and the California Institute of Technology (USA).
%%%%%%%%%%%%%%%%%%%%%%%%%%%%%%%%%%%%%%%%%%%%%%%%%%



\bibliographystyle{mnras}
\bibliography{NuSTAR-IPs.ACCEPTED.bib} % if your bibtex file is called example.bib

%%%%%%%%%%%%%%%%%%%%%%%%%%%%%%%%%%%%%%%%%%%%%%%%%%

%%%%%%%%%%%%%%%%% APPENDICES %%%%%%%%%%%%%%%%%%%%%

%\appendix


%If you want to present additional material which would interrupt the flow of the main paper, it can be placed in an Appendix which appears after the list of references.

%%%%%%%%%%%%%%%%%%%%%%%%%%%%%%%%%%%%%%%%%%%%%%%%%%

% Don't change these lines
\bsp	% typesetting comment
\label{lastpage}
\end{document}